\documentstyle[prb,preprint,aps]{revtex}
\begin{document}
\title{Effect of magnetic and non-magnetic impurities on highly anisotropic
superconductivity.}
\author{A.A. Golubov}
\address{Department of Applied Physics, University of Twente, The Netherlands \\
and Institute for Solid State Physics, Chernogolovka, Russia}
\author{I.I. Mazin}
\address{George Mason University and \\
Complex Systems Theory Branch, Naval Research Laboratory, Washington, DC\\
20375-5320}
\maketitle

\begin{abstract}
We generalize Abrikosov-Gor'kov solution of the problem of weakly coupled
superconductor with impurities on the case of a multiband superconductor
with arbitrary interband order parameter anisotropy, including interband
sign reversal of the order parameter. The solution is given in terms of the
effective (renormalized) coupling matrix and describes not only $T_c$
suppression but also renormalization of the superconducting gap basically at
all temperatures. In many limiting cases we find analytical solutions for
the critical temperature suppression. We illustrate our results by numerical
calculations for two-band model systems.
\end{abstract}

\today 

\section{Introduction}

Recent advance in the field of high-temperature superconductivity, in
particularly discovery of strong anisotropy of the order parameter,
stimulated new interest to the old problem of the effect of (magnetic and
non-magnetic) impurity scattering on superconductivity with high anisotropy.
In a number of theoretical papers published within the last few years
qualitatively new phenomena were uncovered \cite
{muzikar,my,pokrov,kim,abrikos96,maki}. Moreover, detailed experimental
studies of the effect of impurities in high-temperature superconductors are
underway (see e.g. Refs. \onlinecite{Sun,Giap,Bern,Tolpygo} and references therein).

A specific, but representative case of anisotropic superconductivity is
multiband superconductivity (e.g., Ref.%
\onlinecite{my,SMB,Entel,Eilat,Genzel,Kresin}), where the order parameter is
different in different bands. Allen showed in 1978\cite{allen} (see also
Ref. \onlinecite{AM}) that a superconductor with a general anisotropy can be
treated within the same mathematical formalism as a multiband
superconductor, if one expands the order parameter, pairing interaction, and
impurity scattering in terms of the Fermi surface harmonics. In this paper
we derive a general formula, analogous to the Abrikosov-Gor'kov formula for
isotropic superconductors \cite{AG}, but valid for an arbitrary multiband
system. According to the Allen's formalism, this result is easily
generalizable to superconductivity with arbitrary angular anisotropy. We
will also show explicit results for various limiting cases to illustrate the
physics of the interplay between impurity scattering and gap function
anisotropy. We will illustrate the results on a model system with strong
interband anisotropy, namely one where superconductivity in one of two bands
is induced by interband proximity effect.

\section{General theory}

Following the standard way of including the impurity scattering in the BCS
theory\cite{AG}, one writes the equations for the renormalized frequency $%
\tilde{\omega}_n$ and order parameter $\tilde{\Delta}_n$ ($n$ is the
Matsubara index), which completely define the superconductive properties of
the system: 
\begin{mathletters}
\begin{eqnarray}
\hbar \tilde{\omega}_{\alpha n} &=&\hbar \omega _n+\sum_\beta \frac{\hbar ^2%
\tilde{\omega}_{\beta n}}{2Q_{\beta n}}(\gamma _{\alpha \beta }+\gamma
_{\alpha \beta }^s)  \label{AG1} \\
\tilde{\Delta}_{\alpha n} &=&\Delta _\alpha +\sum_\beta \frac{\hbar ^2\tilde{%
\Delta}_{\beta n}}{2Q_{\beta n}}(\gamma _{\alpha \beta }-\gamma _{\alpha
\beta }^s)  \label{AG2} \\
\Delta _\alpha  &=&2\pi T\sum_{\beta ,n}^{0<\omega _n<\omega _D}\Lambda
_{\alpha \beta }\tilde{\Delta}_{\beta n}/{Q}_{\beta n}.  \label{AG3}
\end{eqnarray}
The general form of these equations for strong coupling and general
anisotropy in terms of the Fermi surface harmonics can be found in Ref. 
\onlinecite{AM}
Note that according to Allen's terminology we work in the disjoint
representation, where Fermi surface harmonics are defined separately for
each sheet of the Fermi surface, and take into account only the lowest
harmonic for each sheet. Other notations in Eqs.1 have their usual
meaning: $\omega _n=(2n+1)\pi T$, $Q_{\alpha n}=\sqrt{\tilde{\omega}_{\alpha
n}^2+\tilde{\Delta}_{\alpha n}^2}$, $\gamma _{\alpha \beta }=U_{\alpha \beta
}N_\beta $ is the scattering rate matrix due to nonmagnetic impurities, and $%
\gamma $$_{\alpha \beta }^s=U_{\alpha \beta }^sN_\beta $ is the same for
magnetic impurities. The coupling matrix $\Lambda $ is defined in the same way
as Allen's matrix $\lambda _{\alpha \alpha ^{\prime }}$ \cite{allen}, $%
\Lambda _{\alpha \alpha ^{\prime }}=V_{\alpha \alpha ^{\prime }}^{{\rm %
pairing}}N_{\alpha ^{\prime }}$. Here $N_\alpha $ is the partial density of
states at the Fermi level in the band $\alpha .$ The scattering potential $U$
and the pairing potential $V^{{\rm pairing}}$ are symmetric matrices, while $%
\gamma ,\gamma ^s,$ and $\Lambda $ are not. We shall also introduce the
following useful notations:

\end{mathletters}
\begin{equation}  \label{def1}
\lambda _\alpha =\sum_\beta \Lambda _{\alpha \beta },\,\,\lambda
=\sum_\alpha \lambda _\alpha N_\alpha /N,\,\,N=\sum_\alpha N_\alpha ,
\end{equation}
where $N$ is the total density of states, $\lambda _\alpha $ are partial
electron-phonon coupling constants, which define the electron mass
renormalization in the band $\alpha ,$ and $\lambda $ is the total isotropic
coupling constant, which enters the BCS and Eliashberg equation for
isotropic constant gap superconductivity. Analogously, we shall introduce
partial scattering rates, 
\begin{equation}  \label{def2}
G_\alpha ^{\pm }=\sum_\beta \Gamma _{\alpha \beta }^{\pm },\,\,\,\Gamma
_{\alpha \beta }^{\pm }=\gamma _{\alpha \beta }\pm \gamma _{\alpha \beta
}^s,\,\,g_{\alpha \beta }^{\pm }=\delta _{\alpha \beta }G_\alpha ^{\pm
}-\Gamma _{\alpha \beta }^{\mp }\,\,
\end{equation}

At temperatures close to $T_c$ one can linearize Eqs. 1 with respect to $%
\Delta $. To do so, we introduce, as usual, the renormalization function $%
Z $ and the gap function $\Delta ^{\prime }:$ 
\begin{equation}
Z_{\alpha n}=\tilde{\omega}_{\alpha n}/\omega _n,\,\,\Delta _{\alpha
n}^{\prime }=\tilde{\Delta}_{\alpha n}/Z_{\alpha n}.\,  \label{def3}
\end{equation}

\begin{equation}
Z_{\alpha n}=1+G_\alpha ^{+}/2\omega _n;\,\,\,\,\Delta _{\alpha n}^{\prime
}(1+G_\alpha ^{+}/2\omega _n)=\Delta _\alpha +\sum_{\alpha ^{\prime }}\Delta
_{\alpha ^{\prime }n}^{\prime }\Gamma _{\alpha \alpha ^{\prime
}}^{-}/2\omega _n,  \label{Z}
\end{equation}
which we can solve for $\Delta ^{^{\prime }}:$ 
\begin{equation}
\Delta _{\alpha n}^{\prime }=\sum_{\alpha ^{\prime }}\Delta _{\alpha
^{\prime }}(\delta _{\alpha \alpha ^{\prime }}+g_{\alpha \alpha ^{\prime
}}^{+}/2\omega _n)^{-1}.  \label{Dprime}
\end{equation}
Now from Eq.\ref{AG3} follows that 
\begin{equation}
\Delta _\alpha =2\pi T\sum_{\beta n}^{0<\omega _n<\omega _D}\Lambda _{\alpha
\beta }\omega _n^{-1}\sum_\gamma \Delta _\gamma (\delta _{\gamma \beta
}+g_{\gamma \beta }^{+}/2\omega _n)^{-1}.  \label{sum}
\end{equation}
For weak ($\gamma \ll 2\pi T_c$) and for intermediate ($\gamma \ll \omega _D$%
) scattering the usual trick with subtracting the clean limit, $g=0$, can be
applied, and extending summation to infinity (a useful matrix formula is $(
{\sf \hat {I}+\hat {A}})^{-1}={\sf \hat {I}-\hat {A}(\hat {I}+\hat
{A})}^{-1})$, one gets 
\begin{equation}
\Delta _\alpha =\sum_{\beta \gamma }\Lambda _{\alpha \beta }[L\delta _{\beta
\gamma }-X_{\beta \gamma }]\Delta _\gamma ;\,\,\,X_{\alpha \beta }=2\pi
T_c\sum_n\sum_\gamma (g_{\gamma \beta }^{+}/2)\omega _n^{-1}(\omega _n\delta
_{\alpha \gamma }+g_{\alpha \gamma }^{+}/2)^{-1},  \label{X}
\end{equation}
where $L=\ln (2\gamma ^{*}\omega _D/\pi T_c).\,\,\gamma ^{*}\approx 1.78$ is
the Euler constant. By introducing the eigensystem of $g^{+}$, $g_{\alpha
\beta }^{+}=\sum_\gamma R_{\alpha \gamma }^{-1}d_\gamma R_{\gamma \beta },$
we can express $X$ in terms of the difference between the two incomplete
gamma-functions ($\psi (x)\equiv \sum_{n\geq 0}(n+x)^{-1})$: 
\begin{equation}
\,\,X_{\alpha \beta }=2\pi T_c\sum_\gamma R_{\alpha \gamma
}^{-1}\sum_n\omega _n^{-1}(d_\gamma /2)(\omega _n+d_\gamma /2)^{-1}R_{\gamma
\beta }=\sum_\gamma R_{\alpha \gamma }^{-1}\chi (d_\gamma /4\pi
T_c)R_{\gamma \beta },  \label{defX}
\end{equation}
with $\chi (x)=\psi (1/2)-\psi (1/2+x)$, which is the standard definition of
the matrix function ${\sf \hat{X}=}\chi ({\sf \hat{g}}^{+}{\sf /}4\pi T_c%
{\sf )}.$

This result is analogous to the classic one of Abrikosov and Gor'kov\cite
{AG} (AG), but includes arbitrary anisotropy. Now solving (\ref{X}) for $L$,
we find: 
\begin{equation}
\Delta _\alpha =\sum_\gamma (\Lambda _{\alpha \gamma }^{-1}+X_{\alpha \gamma
})^{-1}L\Delta _\gamma ,  \label{last}
\end{equation}
which means that now $T_c$ is defined by the {\it effective }matrix $\Lambda
_{eff}=(\Lambda ^{-1}+X)^{-1}.$ As it is well known in the multiband
superconductivity theory\cite{ButlerAllen}, in this case $T_c$ is defined by
the usual BCS equation, $T_c=(2\gamma ^{*}\omega _D/\pi )\exp (-1/\lambda
_{\max }),$ where $\lambda _{\max }$ is the maximal eigenvalue of the matrix 
$\Lambda $ (in our case, of the matrix $\Lambda _{eff}).$ As can be seen
immediately from Eqs.(\ref{X}-\ref{last}) and the definition of $g_{\alpha
\beta }$, diagonal nonmagnetic scattering rates $\gamma _{\alpha \beta }$
have dropped out from Eq.\ref{last}. This is the manifestation of the Anderson
theorem for a many band case: intraband scattering does not influence $T_c$
(in the considered Born limit). As will be discussed below, this argument
works only for the intraband non-magnetic scattering, while all other are,
in principle, pair-breaking.

Up to the second order in $\Lambda $ (assuming that $\Lambda X$ is small), 
\begin{equation}
\Lambda _{eff}=\Lambda -\Lambda X\Lambda .  \label{last1}
\end{equation}
If we recall that $\Delta $ forms the eigenvector of $\Lambda $
corresponding to its maximal eigenvalue $\lambda _{eff}$, we can immediately
write the lowest-order correction to $\lambda _{eff}$: 
\begin{equation}
\delta \lambda _{eff}=-\lambda _{eff}^2\sum_{\alpha \beta }\Delta _\alpha
X_{\alpha \beta }\Delta _\beta /\sum_\alpha \Delta _\alpha ^2.
\label{dlambdaeff}
\end{equation}
In the strong scattering case ($\gamma \gg \omega _D)\,$ this formalism
cannot be used. Instead, one should use Eq.\ref{sum} directly.

\section{Critical temperature}

\subsection{Weak scattering}

Let us consider explicitly some interesting limiting cases. For weak
scattering ($\gamma _{\alpha \beta },\gamma _{\alpha \beta }^s\ll T_c)$ one
can use Eq.\ref{dlambdaeff}, and expand $\chi (x\rightarrow 0)=\pi ^2x/2$
and write 
\begin{equation}
\delta T_c/T_c=\delta \lambda _{eff}/\lambda _{eff}^2=-\frac{\sum_{\alpha
\beta }\Delta _\alpha X_{\alpha \beta }\Delta _\beta }{\sum_\alpha \Delta
_\alpha ^2}\approx -\frac{\pi \sum_{\alpha \beta }\Delta _\alpha g_{\alpha
\beta }^{+}\Delta _\beta }{8T_c\sum_\alpha \Delta _\alpha ^2}.  \label{weak}
\end{equation}
When all $\Delta $'s are equal (isotropic case), the standard
Abrikosov-Gor'kov result is recovered: $\delta T_c/T_c=-\pi \sum_{\alpha
\beta }(\Gamma _{\alpha \beta }^{+}-\Gamma _{\alpha \beta
}^{-})/8T_{c0}=-(\pi /4T_{c0})\sum \gamma ^s$, that is, non-magnetic
scattering falls out. On the other hand, in anisotropic case only the
intraband non-magnetic scattering falls out of Eq. \ref{weak}, as for
instance in a two-band case:

\begin{eqnarray}
\delta T_c/T_c &=&-\frac \pi {8T_c}\hat{\Delta}\cdot \left( 
\begin{array}{ll}
2\gamma _{11}^s+\gamma _{12}^s+\gamma _{12} & \gamma _{12}^s-\gamma _{12} \\ 
\gamma _{21}^s-\gamma _{21} & 2\gamma _{22}^s+\gamma _{21}^s+\gamma _{21}
\end{array}
\right) \cdot \hat{\Delta}/\hat{\Delta}\cdot \hat{\Delta}  \label{weak2} \\
&=&-\frac {\pi [\Delta _1^2(2\gamma _{11}^s+\gamma _{12}^s+\gamma
_{12})+\Delta _1\Delta _2(\gamma _{12}^s+\gamma _{21}^s-\gamma _{12}-\gamma
_{21})+\Delta _2^2(2\gamma _{22}^s+\gamma _{21}^s+\gamma _{21})]}{%
8T_c(\Delta _1^2+\Delta _2^2)}.  \nonumber
\end{eqnarray}

The main point of the AG theory\cite{AG} is that $\gamma $$^s$ enters
equations for $\omega $ and $\Delta $ with opposite signs. That is why the
magnetic impurities appear to be pair-breakers, and the non-magnetic ones
not. The above solution shows explicitly that in the multiband case of Eqs.1
only intraband non-magnetic scattering does not influence $T_c$ ($\gamma
_{\alpha \beta }$ drop out). In an interesting limit of two bands, in which
one band is superconducting and another is not, $\lambda _{11}\neq 0,\lambda
_{12}=\lambda _{21}=\lambda _{22}=0$ it follows from Eq. \ref{weak2} that 
\begin{equation}
\delta T_c/T_c=-\frac \pi {8T_c}(2\gamma _{11}^s+\gamma _{12}^s+\gamma
_{12}),  \label{fig1}
\end{equation}
where the first term is the usual AG $T_c$-suppression, and the last two
show that the pair-breaking influence of the non-superconducting band is the
same both for magnetic and nonmagnetic scattering. However, the sign of the
order parameter, induced in the second band, is different: the same for
nonmagnetic and the opposite for magnetic scattering (cf. the $\lambda _2=0$
curves in Fig.\ref{inducedgaps}). Such sign reversal is discussed in more
detail later in the paper.

In the next order in $\lambda _{\beta ,\alpha \neq 1}$ the additional
correction to $\delta T_c/T_c$ is ($\delta T_c/T_c)_1=-\frac \pi {8T_c}%
[(\gamma _{21}^s-\gamma _{21})\lambda _{12}+(\gamma _{12}^s-\gamma
_{12})\lambda _{21}]/(\lambda _{11}-\lambda _{22})$ (this corresponds to the
so-called interband tunneling, specific cases of which are considered in the
literature \cite{Kresin}). In the limit of $\gamma _{ij}^s=0$ the above
expression coincides with that derived in Ref.\onlinecite{Kresin}. Since $%
\gamma _{12}/\gamma _{21}=\lambda _{12}/\lambda _{21}=N_2/N_1$, the last
expression can also be written as 
\begin{equation}
(\delta T_c/T_c)_1=-\frac \pi {4T_c}(\gamma _{12}^s-\gamma _{12})\lambda
_{21}/(\lambda _{11}-\lambda _{22}).  \label{tc1}
\end{equation}
Note that if $\lambda _{21}=0$ suppression of $T_c$ is independent of $%
\lambda _{22},$ as long as $\lambda _{11}>\lambda _{22}.$ It is clearly
seen, for instance, in the left-hand part of Fig.\ref{tcimp}, where the
suppression rate for $\lambda _{21}=0$ and various $\lambda _{22}$ is shown,
and is practically independent on $\lambda _{22}.$ For producing this figure
we have solved Eqs. 1 numerically for two bands, assuming  $\lambda _{21}=$ $%
\lambda _{12}=0,$ $\gamma _{\alpha \beta }^s=\gamma _{11}=\gamma _{22}=0,$
and $\gamma _{12}=\gamma _{21}.$ In full agreement with
Eqs.\ref{fig1},\ref{tc1} $%
T_c$ is first suppressed linearly with the rate $\pi \gamma _{12}/8T_{c0},$
then, at $\gamma _{12}\sim T_c$ it starts to deviate from linearity, and as
it will be proved later in the paper, saturates at some value depending on $%
\lambda _{22}.$

Another important limiting case, also often considered in the literature, is
the limit of the weak anisotropy. Let us assume that $\Delta _\alpha =\bar{%
\Delta}+\delta \Delta _\alpha ,$ where $|\delta \Delta _\alpha |\ll \bar{%
\Delta}.$ The pair-breaking effect of magnetic impurities is then given by
the isotropic AG theory, so it is sufficient to consider only non-magnetic
scattering. Let us also take, for simplicity, an isotropic scattering, $%
g_{\alpha \beta }^{+}=\gamma (\delta _{\alpha \beta }-1)$. Then Eq.\ref{weak}
gives:

\begin{equation}
\delta T_c/T_c=-\frac{\pi \gamma _{tot}}{8T_c}\frac{(\overline{\Delta ^2}-%
\bar{\Delta}^2)}{\ \overline{\Delta ^2}}\approx -\frac{\pi \gamma _{tot}}{%
8T_c}\frac{\overline{\delta \Delta ^2}}{\ \bar{\Delta}^2},  \label{weaka}
\end{equation}
where $\gamma _{tot}$ is the total non-magnetic scattering, summed over all
bands (or Fermi harmonics). Thus in case of weak anisotropy the $T_c$
suppression is given by the AG formula with an effective scattering rate 
$\gamma _{eff}^s=\gamma ^s+(\overline{\delta \Delta ^2}/\bar{\Delta}^2)\gamma$.
This result has often been obtained for the angular gap anisotropy
\cite{hohen,abrikos,pokrov}.

\subsection{Interband sign reversal of the order parameter\label{ISR}}

Returning to the two-band case, we observe that Eq.\ref{last} is invariant
with respect to simultaneous change of signs of $\lambda _{12}$ and $\lambda
_{21}$ and interchange of nondiagonal magnetic and nonmagnetic scattering $%
\gamma _{12}\rightarrow \gamma _{12}^s,$ $\gamma _{21}\rightarrow \gamma
_{21}^s$. This remarkable property is the consequence of the symmetry of the
matrix $\hat{\Lambda}_{eff}$ with respect to the above transformations with
simultaneous reversal of the relative signs of the order parameters $\Delta
_1,\Delta _2.$ One manifestation of this phenomenon is discussed above for
induced superconductivity. Another illustration is given by the symmetric
case $\lambda _{11}=\lambda _{22}\equiv \lambda _{\parallel }$ and $\lambda
_{12}=\lambda _{21}\equiv \lambda _{\perp }$. Then it follows from Eq. \ref
{last} that with sign change of $\lambda _{\perp }$ the role of magnetic and
nonmagnetic interband scattering is completely reversed. Namely, for
positive $\lambda _{\perp }$ ($\Delta _1$ and $\Delta _2$ have the same
signs) only magnetic interband scattering suppresses $T_c$ according to $%
(T_{c0}-T_c)/T_c\approx \pi (\gamma _{12}^s+\gamma _{21}^s)/8T_c$. In the
opposite case of negative $\lambda _{\perp }$ ($\Delta _1$ and $\Delta _2$
have different signs) the magnetic impurities do not influence $T_c$ but the
nonmagnetic ones suppress it according to $(T_{c0}-T_c)/T_c\approx \pi
(\gamma _{12}+\gamma _{21})/8T_c$. The case of arbitrary $\lambda
_{22}/\lambda _{11}$ is shown in Fig.\ref{inducedgaps}, where we show the
numerical solution of Eqs.1 for the same model as we used in Fig.\ref{tcimp}%
: $\lambda _{11}=0.5$ is fixed, and $\lambda _{22}$ changes from 0 to 0.4.
Both magnetic ($\gamma _{12}^s\neq 0,$ $\gamma _{12}=0)$ and non-magnetic ($%
\gamma _{12}^s=0,$ $\gamma _{12}\neq 0)$ impurities are considered. In the
first band the order parameter is suppressed equally by magnetic and
nonmagnetic impurities: the solid curves in Fig.\ref{inducedgaps} are the
same for both cases. The order parameter in the second band has the same
absolute value for pure magnetic or for pure nonmagnetic scattering, but its
sign is different in the two cases. Moreover, even if $\lambda _{12},\lambda
_{21}\neq 0,$ but $\lambda _{12},\lambda _{21}\ll \lambda _{11},\lambda _{22}
$, there still is a possibility of the interband sign reversal of the gap
due to magnetic impurities. This happens when nondiagonal elements in the
effective $\Lambda $ matrix in Eq.\ref{last} become negative: $\lambda
_{12}^{eff}=\lambda _{12}+\pi \lambda _{11}\lambda _{22}(\gamma _{12}-\gamma
_{12}^s)/8T_{c0},$ which does happen if $\gamma _{12}^s$ is sufficiently
large. Then the order parameters in different bands have different signs,
i.e. solution with sign$(\Delta _\beta )=-$sign$(\Delta _\alpha )$
corresponds to a minimum energy. This sign reversal leads to an interesting
effect: if one starts from a pure superconductor with weak interband
coupling, and suppresses $T_c$ by adding interband magnetic impurity
scattering, at some critical scattering strength suppression rate drops
drastically. The final comment to Fig.\ref{inducedgaps} is that it shows
either solely magnetic, or solely nonmagnetic scattering. When both kinds of
scattering are present, the order parameter in the second band is much
smaller than in the either pure case, and becomes zero when magnetic and
nonmagnetic scattering are equally strong. Numerical illustration of this
effect can be found in Ref. 
\onlinecite{my}


This situation is closely analogous to the known case of {\it d}-pairing,
where isotropic non-magnetic impurity scattering leads to an AG $T_c$
suppression, but with a factor of two smaller coefficient (cf. Refs.%
\onlinecite{millis,radtke}). If we label those parts of the Fermi surface
that have positive order parameter as 1 and those which have negative order
parameter as 2, then only in the ``interband channel'' the nonmagnetic
impurities are pair-breaking, while the magnetic impurities are
pair-breaking only in the ``intraband channel''. Correspondingly, the
effective pair-breaking scattering rate will be $\gamma _{11}^s+\gamma
_{22}^s+\gamma _{12}+\gamma _{21}=(\gamma _{tot}+\gamma _{tot}^s)/2.$ Note
that isotropic magnetic scattering results 
in exactly as much pair-breaking in terms
of $T_c$ as isotropic non-magnetic scattering, contrary to the popular
misconception that only nonmagnetic impurities are suppressing $T_c$ in $d-$%
wave superconductors. Interestingly, if isotropic magnetic and non-magnetic
scattering are both present, and have equal strength, $T_c$ suppression rate
is the same for the $s-$ and $d-$wave superconductors. If only magnetic
scattering is present, $T_c$ is suppressed twice faster in an $s$%
-superconductor. In fact, most of these statements are not specific for the $%
d$-pairing, but are true for any superconductor with zero average order
parameter and non-zero average square for the order parameter. Let us, for
example, prove that in such a superconductor isotropic magnetic and
non-magnetic scatterings both have the same effect on $T_c.$ According to
Eq. \ref{weak}, $T_c$ suppression rate is proportional to 
\[
\langle \Delta _{{\bf k}}g_{{\bf kk}^{\prime }}^{+}\Delta _{{\bf k}}\rangle
_{{\bf k,k}^{\prime }}=\langle \Delta _{{\bf k}}^2G_{{\bf k}}^{+}\rangle _{%
{\bf k}}-\langle \Delta _{{\bf k}}\Gamma _{{\bf kk}^{\prime }}^{-}\Delta _{%
{\bf k}^{\prime }}\rangle _{{\bf kk}^{\prime }}=\langle \Delta _{{\bf k}%
}^2\langle \Gamma _{{\bf kk}^{\prime }}^{+}\rangle _{{\bf k}^{\prime
}}\rangle _{{\bf k}}-\langle \Delta _{{\bf k}}\Gamma _{{\bf kk}^{\prime
}}^{-}\Delta _{{\bf k}^{\prime }}\rangle _{{\bf kk}^{\prime }}, 
\]
where we used {\bf k} and ${\bf k}^{\prime }$ for indices to emphasize that
the formalism is valid both for interband or for angular anisotropy. For
isotropic scattering, $\gamma _{{\bf k,k}^{\prime }}=\gamma ,$ $\gamma _{%
{\bf k,k}^{\prime }}^s=\gamma ^s,$ this equation reduces to

\[
\langle \Delta _{{\bf k}}^2\rangle (\gamma +\gamma ^s)-\langle \Delta _{{\bf %
k}}\Delta _{{\bf k}^{\prime }}\rangle (\gamma -\gamma ^s)=(\langle \Delta _{%
{\bf k}}^2\rangle +\langle \Delta _{{\bf k}}\Delta _{{\bf k}^{\prime
}}\rangle )\gamma ^s+(\langle \Delta _{{\bf k}}^2\rangle -\langle \Delta _{%
{\bf k}}\Delta _{{\bf k}^{\prime }}\rangle )\gamma . 
\]
For isotropic $s$-wave superconductors, $\langle \Delta _{{\bf k}}^2\rangle
=\langle \Delta _{{\bf k}}\Delta _{{\bf k}^{\prime }}\rangle =\Delta ^2,$
and the $T_c$ suppression rate does not depend on $\gamma .$ For a
superconductor where $\langle \Delta _{{\bf k}}\Delta _{{\bf k}^{\prime
}}\rangle =0,$ $\langle \Delta _{{\bf k}}^2\rangle \neq 0,$ a specific case
of which is a $d$-wave superconductor, the suppression rate is proportional
to $(\gamma +\gamma ^s),$ as we have conjectured before.

\subsection{Strong scattering}

Let us now go beyond the weak scattering limit, so that we cannot any more
use the expansion in $X\Lambda $ in Eq.\ref{last}. In accordance with the AG
result, the critical temperature vanishes at some finite rate of intraband
magnetic scattering $\gamma _{\alpha \beta }^s\sim T_{c0}$. The situation is
qualitatively different with respect to interband scattering. We will show
that in the strongly anisotropic case of $\lambda _{11},\lambda _{22}\gg
\lambda _{\alpha \neq \beta }$ the critical temperature does not vanish even
in the regime of very strong interband scattering. Let us first consider
intermediate scattering regime $\pi T_c\ll \gamma _{\alpha \beta }\ll \omega
_D.$ In this case one still can use Eqs.(\ref{X}-\ref{last}). Using
expansion 
\[
\chi (x\rightarrow \infty )=\log (4\gamma ^{*}x+const) 
\]
we obtain that 
\[
\Lambda _{eff}=\Lambda -\Lambda \cdot \log (\gamma ^{*}{\sf \hat g^+}/\pi
T+const)\cdot \Lambda , 
\]
which has a particularly simple form for the case we are interested in, $%
\gamma _{11}=\gamma _{22}=0,$ ${\sf \hat g^+}=\left( 
\begin{array}{cc}
\gamma _{12} & -\gamma _{12} \\ 
-\gamma _{21} & \gamma _{21}
\end{array}
\right) :$%
\[
\Lambda _{eff}=\Lambda -(\gamma _{12}+\gamma _{21})^{-1}\log [4\gamma
^{*}(\gamma _{12}+\gamma _{21})/\pi T_c]\Lambda \cdot {\sf \hat g^+}%
\cdot \Lambda . 
\]

\begin{equation}
\frac{T_c}{T_{c0}}=\left[ \frac{\pi T_c}{2\gamma ^{*}(\gamma _{12}+\gamma
_{21})}\right] ^{\gamma _{12}/(\gamma _{12}+\gamma _{21})},  \label{tc-inter}
\end{equation}
where we assumed, to be specific, that $\lambda _{11}\geq \lambda _{22}.$
Solving for $T_c,$ we get 
\begin{equation}
\frac{T_c}{T_{c0}}=\left[ \frac{\pi T_{c0}}{2\gamma ^{*}(\gamma _{12}+\gamma
_{21})}\right] ^{\gamma _{12}/\gamma _{21}}=\left[ \frac{\pi T_{c0}}{2\gamma
^{*}(\gamma _{12}+\gamma _{21})}\right] ^{N_2/N_1},  \label{tc-inter-1}
\end{equation}
where $N_{1,2}$ are the densities of states in the two
bands (the last equality appears because $\gamma _{\alpha \beta }/\gamma
_{\beta \alpha }=N_\beta /N_\alpha )$.

Eq. \ref{tc-inter-1} gives the limit $T_c\rightarrow 0$ when $\gamma
_{\alpha \beta }\rightarrow \infty .$ However, Eq.\ref{last} becomes invalid
in this regime, namely when the interband scattering rate $\gamma _{\alpha
\beta }$ exceeds the characteristic electronic energy scale $\omega _D$
which is relevant for the Cooper pairing. In this case, we have to go back
to Eq. \ref{sum}. This equation can be solved analytically in an important
regime of the isotropic superstrong interband scattering, $\gamma _{\alpha
\beta }=\gamma (1-\delta _{\alpha \beta })N_\beta $. In this regime, $%
g_{\alpha \beta }^{+}=\gamma (\delta _{\alpha \beta }N-N_\beta ),$ where $%
N=\sum_\alpha N_\alpha .$ To handle Eq. \ref{sum} we first need to transform
the matrix 
\[
(2\omega _n\delta _{\alpha \beta }+g_{\alpha \beta }^{+})^{-1}=[(2\omega
_n+\gamma N)\delta _{\alpha \beta }-\gamma N_\beta ]^{-1}=(2\omega _n+\gamma
N)^{-1}[\delta _{\alpha \beta }-\gamma N_\beta /(2\omega _n+\gamma N)]^{-1}
\]
to a more tractable form. Expanding the square bracket in series in $\gamma
N_\beta /(2\omega _n+\gamma N)$ and collecting the appropriate terms, we
observe that 
\[
(2\omega _n\delta _{\alpha \beta }+g_{\alpha \beta }^{+})^{-1}=(2\omega
_n+\gamma N)^{-1}(\delta _{\alpha \beta }+\gamma N_\alpha /2\omega _n),
\]
which in the sought limit $\gamma
\rightarrow \infty $ is simply 
$ N_\alpha/2N\omega _n.$ 
Thus 
\begin{equation}
\Delta _\alpha =\sum_\beta \Lambda _{\alpha \beta }\sum_k\frac{%
N_k\Delta _k}N\sum_n\frac{2\pi T}{\omega _n}  \label{tc-strong}
\end{equation}
which has the solution 
\begin{equation}
\Delta _\alpha =\lambda _\alpha \bar{\Delta}\sum_n\frac{2\pi T}{\omega _n}
\label{del-strong}
\end{equation}
where $\lambda _\alpha =\sum \Lambda _{\alpha \beta }$ is the mass
renormalization parameter, and the average gap $\bar{\Delta}=\sum_\alpha
N_\alpha \Delta _\alpha /N$ satisfies the regular BCS equation with the
isotropically averaged coupling: 
\begin{equation}
\overline{\Delta }=\lambda \overline{\,\,\Delta }\,\,\sum_n\frac{2\pi T}{%
\omega _n}\,,\,\,\,\,\,\,\,\,\lambda =\sum_\alpha N_\alpha \lambda _\alpha
/N.  \label{lam-strong}
\end{equation}
So, in the superstrong coupling regime $T_c$ saturates at a limiting value,
which is actually the critical temperature calculated in fully isotropic BCS
theory. This regime corresponds to the so-called Cooper limit investigated
previously for the proximity-effect coupled systems \cite{Kresin,mcm}. Note
that the order parameters in the individual bands are nevertheless
different, specifically, $\Delta _\alpha =\bar{\Delta}\lambda _\alpha
/\lambda .$ This does not mean that the observable zero-temperature gaps are
going to be different. In fact, they are the same, as discussed in the next
Section and illustrated on Fig. \ref{DOS}.

We illustrate the above discussion of the $T_c$ suppression by a numerical
solution of the Eqs.\ref{sum} for a two-band case with strong interband
anisotropy and $\lambda _{12},\ \lambda _{21}\ll \lambda _{11},\lambda _{22}$, 
$\gamma _{21}^s=\gamma _{12}^s$, corresponding to equal densities of states
in the two bands $N_1=N_2$. We have chosen $\lambda _1=0.2$, $\lambda
_{12},\ \lambda _{21}=0$ and $\gamma _{21}^s=\gamma _{12}^s=0$. The results
of calculations of $T_c$ vs $\gamma _{12}$ are shown in Fig.\ref{tcimp} for
various values of $\lambda _2$. In accordance with the above analytical
results, $T_c$ first drops steeply as $\gamma _{12}$ increases and then
saturates at some finite value when $\gamma _{12}\sim \omega _D.$ The
saturation value depends on $\lambda _2$ in accordance with Eq.\ref
{lam-strong}. The suppression of $T_c$ remains the same when nonmagnetic
scattering is zero, $\gamma _{12}=0$, but $\gamma _{12}^s$ is finite, except
that the order parameter $\Delta _{2,n=0}$ has the sign opposite to that of $%
\Delta _{1,n=0}$, as discussed above and illustrated on Fig.\ref{inducedgaps}.
Both kinds of impurities, $\gamma _{12}^s$ and $\gamma _{12}$, suppress
the critical temperature in this case according to Eqs.\ref{tc1}-\ref
{lam-strong}.

\section{Density of states and superconducting gap}

The discussion of the critical temperature suppression was based on the
solutions of the linearized equations. To obtain the density of states, the
nonlinear Eqs.1 should be solved. In the presence of impurities there is no
distinct gap, in the sense that the minimal excitation energy does not
coincide with the maximum in the density of states. The latter is defined in
terms of $\Delta _{n\alpha }^{\prime }.$ Namely, the superconducting density
of states $N(\omega )$ in units of the normal density of states at the Fermi
level $N_0$ is 
\[
\frac{N(\omega )}{N_0}={Re}\frac \omega {\sqrt{\omega ^2-\Delta ^{\prime
2}(\omega )}}, 
\]
where $\Delta ^{\prime }(\omega )$ is the analytical continuation of $\Delta
_n^{\prime }.$ An analytical solution for $\Delta _{n\alpha }^{\prime }$ is
not straightforward to obtain; however, some properties of the numerically
obtained solutions for $\Delta _{n\alpha }^{\prime }$ are already
illustrated above in Fig.\ref{inducedgaps}. Moreover, there are some
rigorous statements that can be made about $\Delta _{n\alpha }^{\prime }$.

Let us consider again the limit of the superstrong isotropic non-magnetic
scattering. We have shown above that $T_c$ in this case is reduced to $T_c$
of the equivalent BCS system with the isotropic coupling constant. The same
statement appears to be true for the gap in the excitation spectrum. Indeed,
following AG, we can define $u_{\alpha n}=\omega _n/\Delta _{in}^{\prime },$
and, after the usual transformation $\omega _n\rightarrow -i\omega ,$ $%
u_{\alpha n}\rightarrow -iu_\alpha ,$ $\Delta _{\alpha n}^{\prime
}\rightarrow \Delta _\alpha ^{\prime }(\omega ),$ we can write down the
multiband analog of the Eq.42$^\prime $ of AG: 
\begin{equation}
\frac \omega {\Delta _\alpha }=u_\alpha -\frac 1{\Delta _\alpha }\sum_\beta 
\frac{u_\alpha \Gamma _{\alpha \beta }^{+}-\Gamma _{\alpha \beta
}^{-}u_\beta }{\sqrt{1-u_\beta ^2}}.  \label{AG45}
\end{equation}
In the absence of magnetic impurities we let $\Gamma _{\alpha \beta
}^{+}=\Gamma _{\alpha \beta }^{-}=\gamma ,$ and tend $\gamma \rightarrow
\infty .$ Evidently, a solution of Eq. \ref{AG45} in this limit exists only
if $u_\alpha =u_\beta ,$ and correspondingly $\Delta _\alpha ^{\prime }$ do
not depend on $\alpha .$ We conclude that in this limit the reduced density
of states is the same in all bands, and in fact coincides with that of the
isotropic BCS model with the gap determined from the nonlinear analog of Eq.%
\ref{lam-strong}: $\overline{\Delta }=\lambda \overline{\Delta }\sum_n2\pi T/%
\sqrt{\omega _n^2+\overline{\Delta }^2}\,,\,$ with$\,\,\,\lambda
=\sum_\alpha N_\alpha \lambda _\alpha /N.$

The evolution of the densities of states in a multiband non-magnetic
scattering case is shown on Fig.\ref{DOS}. Here, we show the results of
numerical solution of Eqs. 1 in the weak coupling regime with $\lambda _1=0.5
$, $\lambda _{12},\ \lambda _{21},\lambda _{22}=0$ and $\gamma _{\alpha
\beta }^s=0$. Only nonmagnetic interband scattering $\gamma _{12}=\gamma
_{21}$ is included. In the clean limit, the two bands show two different
excitation gaps. In accordance with earlier calculations\cite{mcm,Schopohl},
any weak, but finite impurity scattering mixes the pairs in the two bands,
so that the first band (with the larger gap, {\it i.e. }more
superconducting) develops a tail in the density of states which extends all
the way down to the second-band gap. Except for this tail, which consists of
the normal excitations of the second band, scattered into the first band by
impurities, the density of states still looks similar to the clean-limit one.
Upon the increase of the scattering rate, the low-energy tail in the first
band density of states grows, and the minimal gap, the gap in the second
band grows as well. This reflects the fact that larger number of pairs is
scattered into the second band and induced the interband-proximity-effect
superconductivity there. Thus the {\it decrease} in the critical temperature
of the system is accompanied by the {\it increase} of the minimal gap in the
excitation spectrum.

Next, let us include interband magnetic scattering into Eq.\ref{AG45}. Then
in the considered regime $\gamma \rightarrow \infty $ we have $\Gamma
_{\alpha \beta }^{+}=\gamma +\gamma ^s,\Gamma _{\alpha \beta }^{-}=\gamma
-\gamma ^s$ and $u_\alpha =u_\beta \equiv u$ (we assumed $\gamma _{\alpha
\neq \beta }^s\equiv \gamma ^s$). As a result, the densities of states in
each band are given by $N(\omega )=\mathop{\rm Re}u/\sqrt{u^2-1}$, where $u$
is a solution of the equation 
\begin{equation}
\frac \omega \Delta =u(1-\frac{2\gamma ^s}\Delta \frac 1{\sqrt{1-u^2}}).
\label{AG-magn}
\end{equation}
An energy gap corresponds to maximum real solution for $u$ in the interval $%
u<1$ and the pair-breaking rate is given by $2\gamma ^s$. Thus, with
increase of non-magnetic scattering we have a crossover from the state with
different signs of order parameters in different bands (for zero $\gamma $)
to the isotropic state (for $\gamma \rightarrow \infty $). This isotropic
state may be normal, gapless or gapped, depending on the value of $2\gamma ^s
$. Following the AG analysis, we obtain that an energy gap at $\gamma
\rightarrow \infty $ will exist if $\gamma ^s<\exp (-\pi /4)\Delta _0/2$,
where $\Delta _0$ is the BCS gap at $T=0$. This case is particularly
interesting: since in the isotropic ($\gamma \rightarrow \infty $) limit
there is a finite gap and a finite positive order parameter in both bands,
and in the opposite limit of small $\gamma $ the order parameter in one band
is negative, it is clear that at intermediate values of interband scattering 
$\gamma $ a gapless state should be crossed over. The last statement is in
agreement with the result of Ref.\onlinecite{muzikar} that for an order
parameter with a sign change and a nonzero Fermi-surface average, a gapless
state develops with an increase of impurity concentration, but the gap is
restored at large concentration of impurities. To illustrate such
crossover, induced
by magnetic scattering, we show in Fig.\ref{dos-magn} the results
of our numerical calculations for a weak coupling 2-band model ($\lambda
_1=0.5,\ \lambda _{12},\ \lambda _{21},\lambda _{22}=0$ and $\gamma
_{12}^s=\gamma _{21}^s=T_c/2$. The total density of states is shown at
different values of interband nonmagnetic scattering rate $\gamma $.
According to the discussion in section \ref{ISR}, for small $\gamma$
both order parameters have different signs.
In this case $\gamma ^s$ has no pair-breaking effect, and small gap (negative 
order parameter) is induced in the second band by the magnetic
scattering. On the contrary, the interband $\gamma$ is in
this situation pair-breaking. 
The shape of the density of states shows two characteristic
peaks, one, at about 0.1$\Delta _0,$ due to this induced energy gap in the
second band, and another just below $\Delta _0$ from the gap in the first
band.

 With the increase of the nonmagnetic scattering rate the order
parameter in the second band becomes smaller in the absolute value, still
remaining negative. The lower peak in the density of states gets washed out
and the minimal energy gap becomes smaller. When $\gamma $ approaches $%
\gamma ^s$ this small gap vanishes, although there is still a
distinguishable peak in the density of states coming from the gap in the
first band.
At larger $\gamma >>\gamma ^s$ both gaps have again the same sign and 
now  it is $\gamma ^s$, which is pair-breaking.
As one again can see from Fig.\ref{dos-magn},
a small gap is restored for the last two curves, corresponding to 
$\gamma=2 \gamma ^s=T_c,$ and $\gamma=20 \gamma ^s=10T_c$.
Note that at $\gamma >> T_c ,$ the gap 
cannot any more be ascribable to any of the two bands, but corresponds to a
fully isotropic superconductivity, as described by Eqs.\ref{tc1}-\ref
{lam-strong}.

\section{Conclusions}

In conclusion, we generalize the Abrikosov-Gor'kov solution on the case of
arbitrary interband anisotropy of the pairing interaction, and arbitrary
strength and anisotropy of magnetic and/or nonmagnetic impurity scattering.
The results are illustrated on model two-band systems with interband
anisotropy and various kinds of impurity scattering. In case of weak
scattering, we found an analytic solution, analogous to the isotropic
solution of Abrikosov-Gor'kov. For weak anisotropy, this solution yields the
critical temperature suppression proportional to the mean square variation
of the order parameter, the fact earlier pointed out by several authors in
various special cases. We also prove analytically that the superconductivity
suppression by isotropic magnetic and isotropic non-magnetic impurities is
exactly the same when the average order parameter is zero ({\it e.g.}, in
case of $d$-pairing). We also give an analytical solution for $T_c$ in the
two-band model in case of intermediate-strength scattering. In case of
superstrong scattering we find a solution for $T_c$ for arbitrary
anisotropy. We also discuss the evolution of the density of states with the
increase of the impurities concentration (or scattering strength).

{\it Acknowledgments. }We acknowledge useful discussions with O.V. Dolgov,
V.Z. Kresin, S.A. Wolf, W.E. Pickett, 
S.V. Shulga and S. Tolpygo. This research was in part supported by
the INTAS grant N 94-2452 (AAG) and by the Office of Naval Research (IIM).

\figure{\ 
\caption{$T_c$ suppression by interband scattering.
Dots on the right axis show the asymptotic value of $T_c$ at
$\gamma \rightarrow \infty$, according to Eq. \ref{lam-strong}.
} \label{tcimp} }

\figure{\ 
\caption{Dependence of the order parameters
$\Delta_{1,n=0}$ and $\Delta_{2,n=0}$
 in two bands in a two-band
model with $\lambda_{12}= \lambda_{21}=0$ and different ratios 
$\lambda_{22}/\lambda_{11}$, on the interband impurity scattering.
Solid lines show the order parameter in the first (``superconducting''),
dashed and dotted lines show the order parameter in the second band, where
superconductivity is induced by impurity scattering. Dashed lines correspond to
non-magnetic interband scattering, dotted lines to magnetic interband
scattering.
} \label{inducedgaps} }

\figure{\ 
\caption{Superconducting density of states in a two-band model with 
$\lambda_{12}= \lambda_{21}=\lambda_{22}=0$, $\lambda_{11}=0.5$. Only
interband non-magnetic scattering is included. The solid lines show DOS in the
band 1 and the dashed lines in the band 2. Note that both DOS 
coincide in the regime of strong scattering.
} \label{DOS} }

\figure{\ 
\caption{Total superconducting density of states in a two-band model with
$\lambda_{12}= \lambda_{21}=\lambda_{22}=0$, $\lambda_{11}=0.5$.} \label
{dos-magn} }

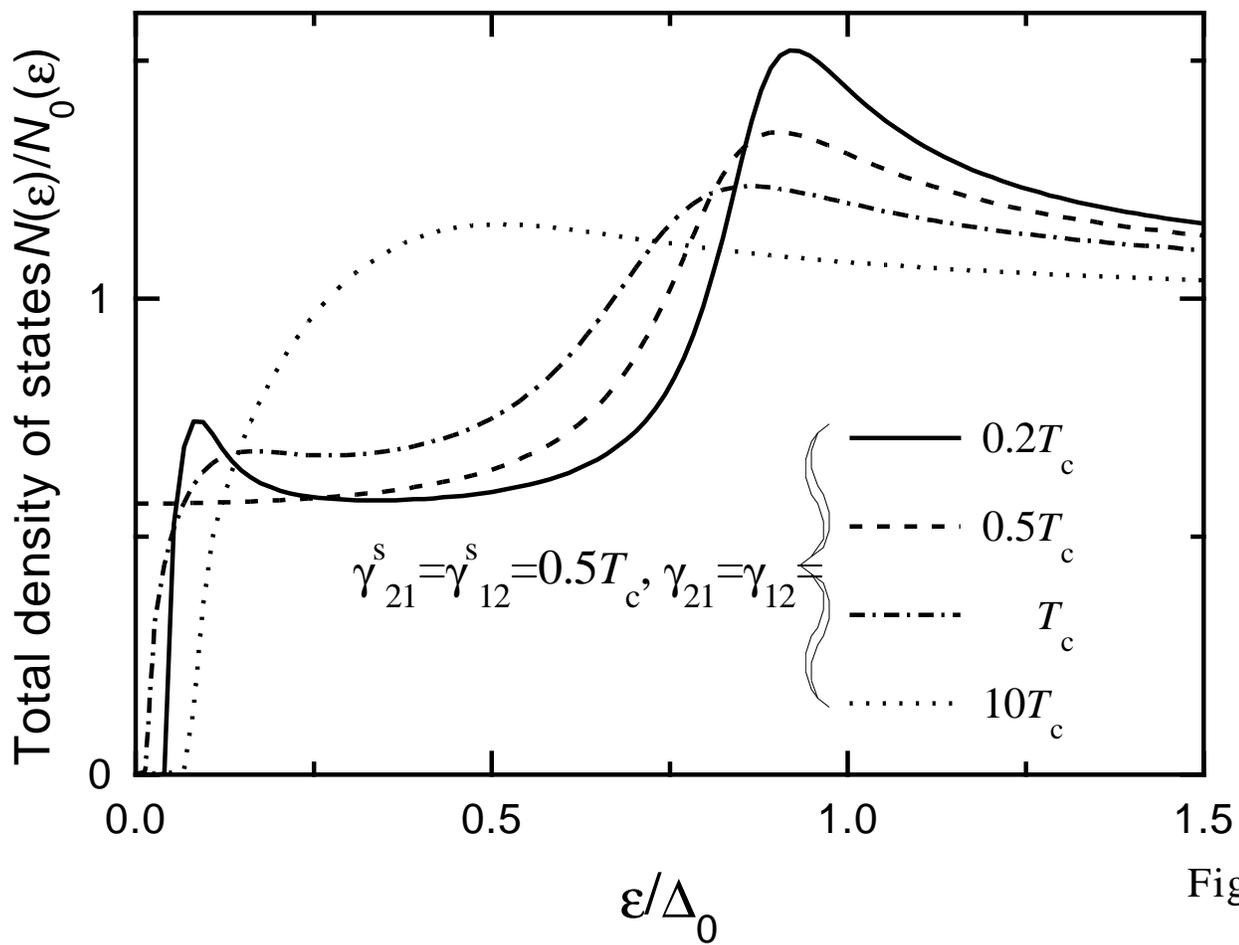
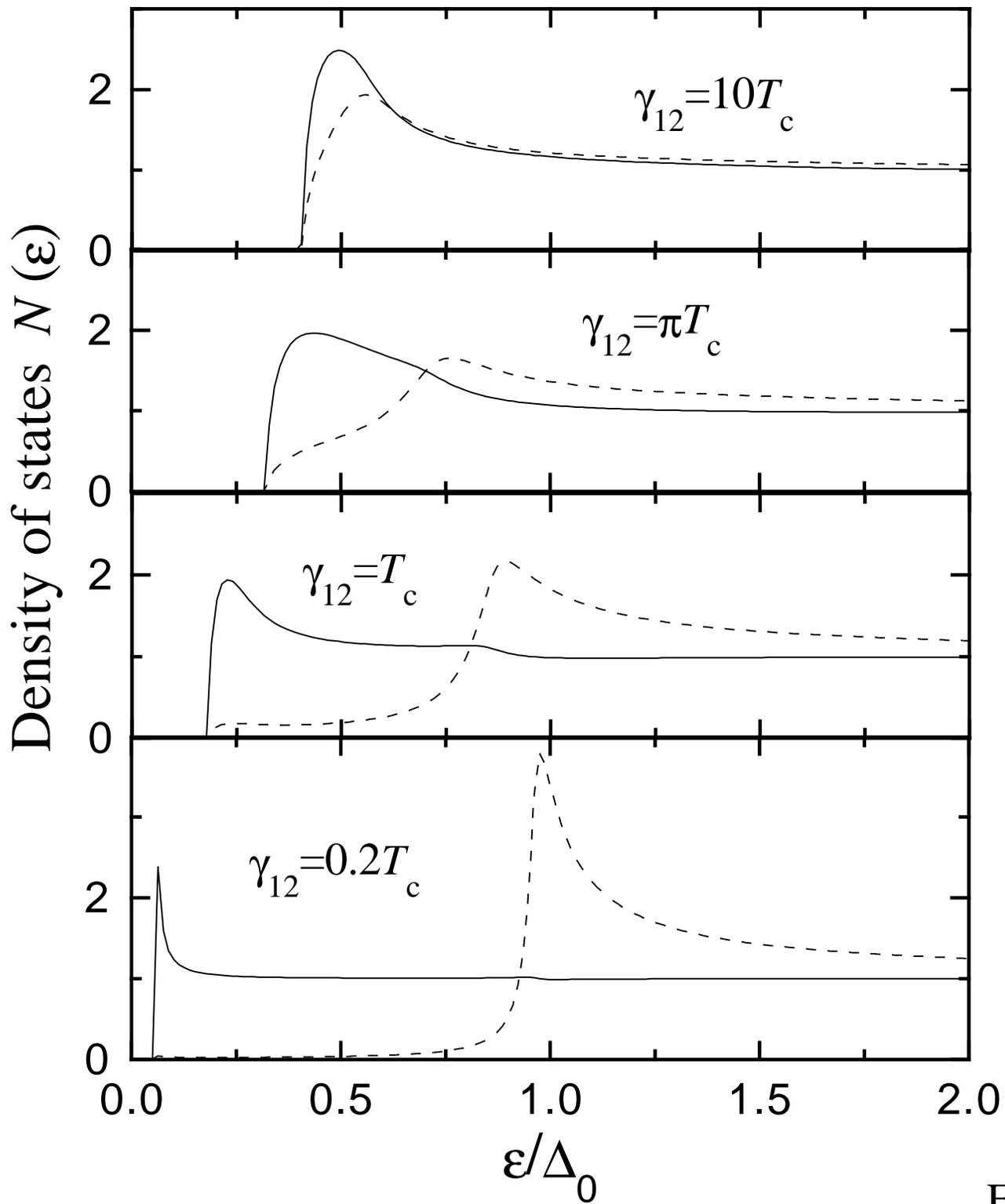
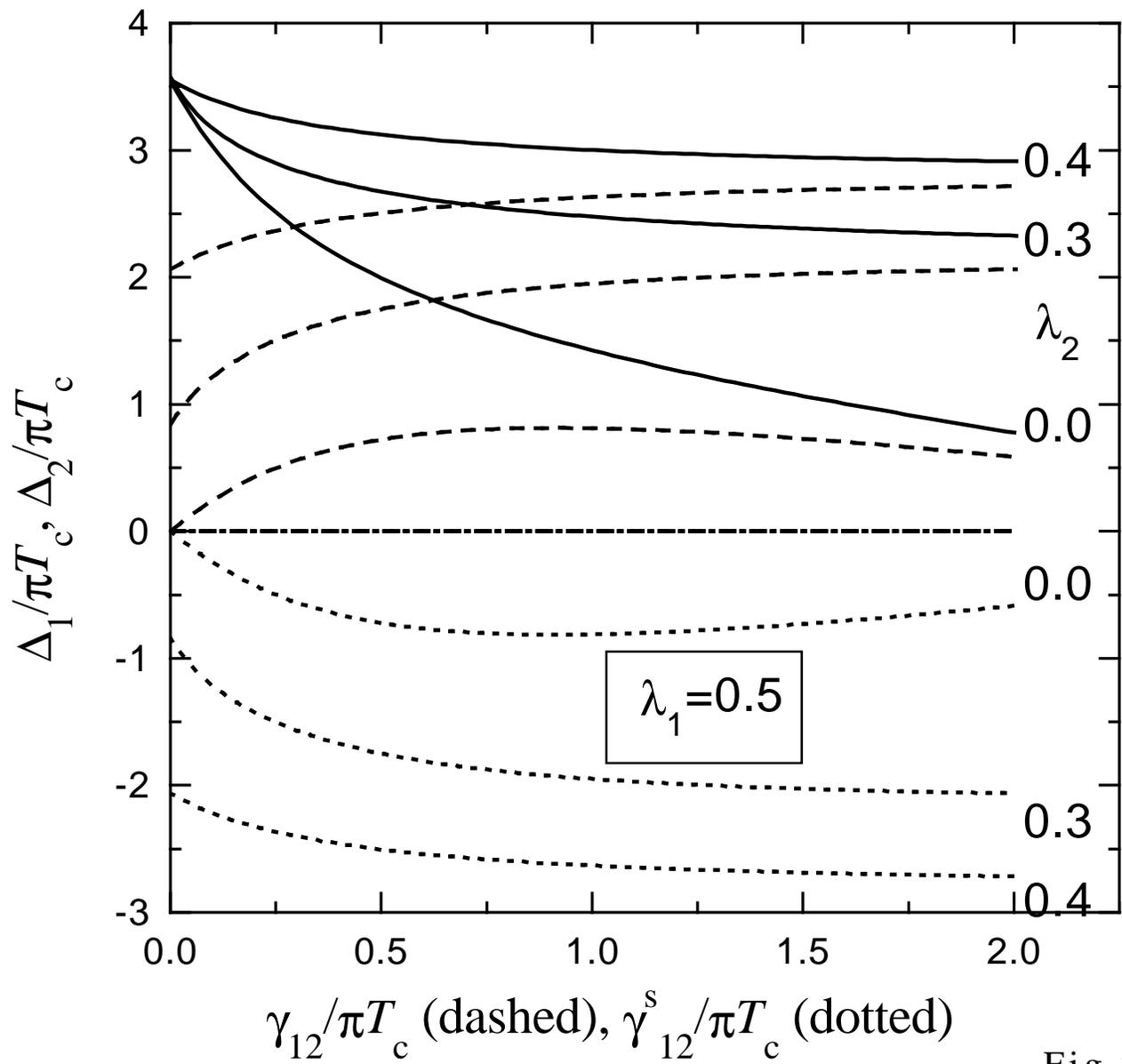
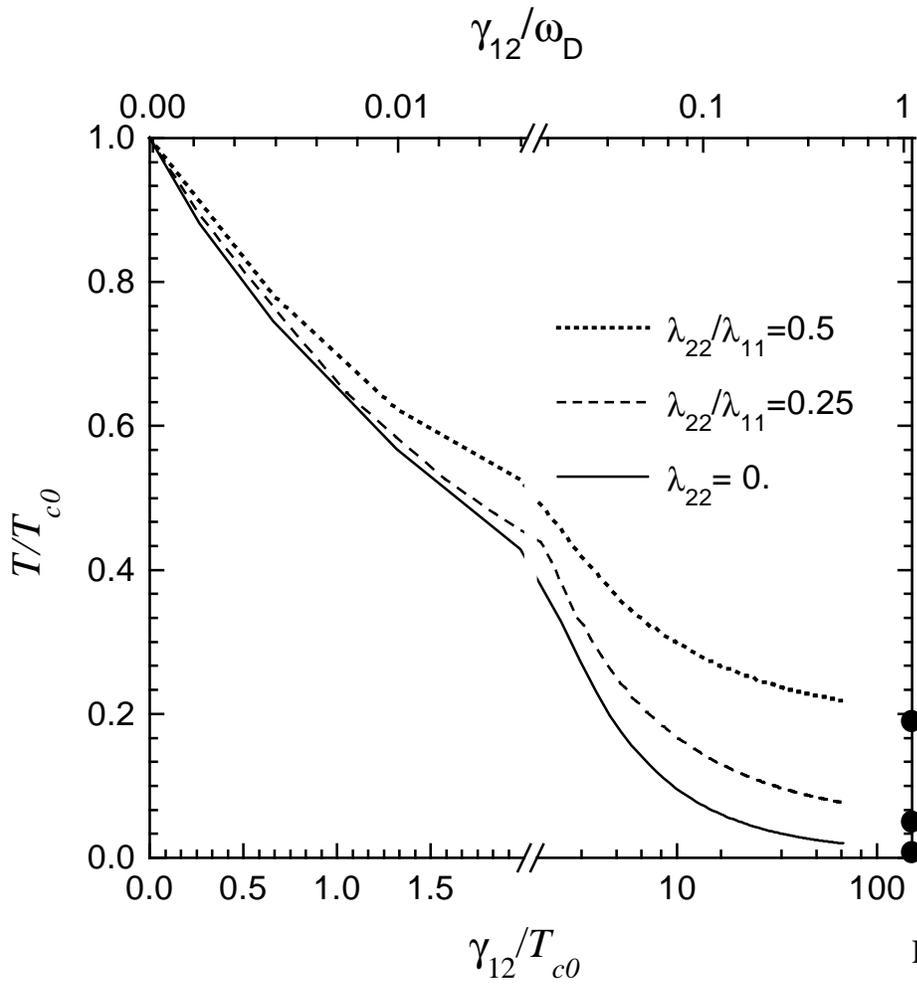
\end{document}